\documentclass[%
 aip,
 amsmath,amssymb,
 reprint,%
]{revtex4-1}

\usepackage{graphicx}
\usepackage{dcolumn}
\usepackage{bm}

\usepackage[utf8]{inputenc}
\usepackage[T1]{fontenc}
\usepackage{mathptmx}
\usepackage{color}

\begin{document}

\preprint{AIP/123-QED}

\title{Translationally Invariant Generation of Annular Beams using Thin Films}

\author{Naresh Sharma}
 \affiliation{Department of Electrical Engineering, Indian Institute of Technology Kanpur, Kanpur-208016, UP, India.}
\author{Govind Kumar}%
\affiliation{The Centre for Lasers and Photonics, Indian Institute of Technology Kanpur, Kanpur-208016, UP, India}%

\author{Vivek Garg}
\affiliation{Department of Mechanical Engineering, Indian Institute of Technology  Bombay, Mumbai-400076, Maharashtra, India}%

\author{Rakesh G. Mote}
\affiliation{Department of Mechanical Engineering, Indian Institute of Technology  Bombay, Mumbai-400076, Maharashtra, India}%

\author{R. Vijaya}
\affiliation{The Centre for Lasers and Photonics, Indian Institute of Technology Kanpur, Kanpur-208016, UP, India}%
\affiliation{Department of Physics, Indian Institute of Technology Kanpur, Kanpur-208016, UP, India}

\author{Shilpi Gupta}
 \affiliation{Department of Electrical Engineering, Indian Institute of Technology Kanpur, Kanpur-208016, UP, India.}
 \email{ShilpiG@iitk.ac.in}

\date{\today}

\begin{abstract}
Thin film optical elements exhibiting translational invariance, and thus robustness to optical misalignment, are crucial for rapid development of compact and integrated optical devices. In this letter, we experimentally demonstrate a beam-shaping element that generates an annular beam by spatially filtering the fundamental Gaussian mode of a laser beam. The element comprises of a one-dimensional photonic crystal cavity fabricated using sputtered thin films. The planar architecture and in-plane symmetry of the element render our beam-shaping technique translationally invariant. The generated annular beam is sensitive to the polarization direction and the wavelength of the incident laser beam. Using this property of the annular beam, we show simultaneous generation of concentric annular beams of different wavelengths. Our experimental observations show an excellent agreement with simulation results performed using finite-difference time-domain method. Such a beam-shaping element has applications in areas ranging from microscopy and medicine to semiconductor lithography and manufacturing in microelectronics industry.  
\end{abstract}

\maketitle

\section{\label{sec:level1}Introduction}

\begin{figure}[t]
\centering\includegraphics[width=8.5 cm]{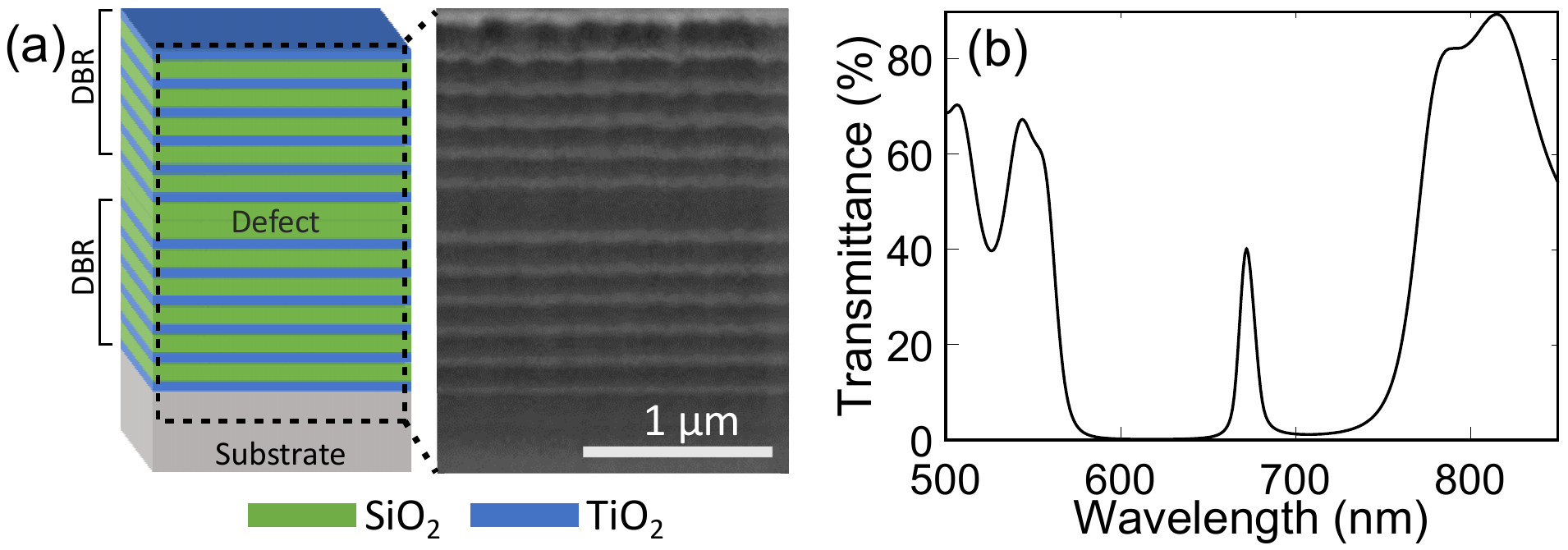}
\caption{\textbf{Characterization of photonic crystal cavity}: (a) A schematic and a scanning electron microscope image of a one-dimensional photonic crystal cavity. DBR: Distributed Bragg Reflector. (b) Transmission spectrum of the photonic crystal cavity.} 
\vspace{-1.5em}
\label{Fig:Fig1_SEMandTransmission}
\end{figure} 

Beam-shaping techniques transform the spatial profile of a radiation beam by changing either the intensity profile or the phase profile or both \cite{Dickey2006}. Among many beam shapes that the fundamental Gaussian mode of a laser beam has been transformed into \cite{Duocastella2012, Li2002,
Liu2007, klas2018}, annular beam is a versatile beam shape that finds applications in a variety of fields including optical microscopy \cite{Sick2000,Klar2000}, optical trapping \cite{Woerdemann2013,Rodrigo2004,lenton2019}, 
laser micromachining and photopolymerization \cite{Duocastella2012,Winfield2007}, and medicine \cite{Ren1990}. Annular beams are also employed for generating Bessel beams that are useful for quantum and optical communication \cite{Chu2015}. 

Traditionally, annular beams are generated in free space using spatial light modulators \cite{lenton2019}, drilled mirrors \cite{klas2018}, shading masks \cite{kitamura2010}, and axicons \cite{livneh2018, wang2015,Winfield2007}. Additionally, more sophisticated refractive elements such as holographic filters \cite{Turunen1988}, diffractive ring lenses \cite{Xin2014}, and metasurfaces \cite{Chen2017} have been employed for generating higher order Bessel beams that are non-diffracting annular beams. Most of these optical elements are optimized for generating a single annular beam, are difficult to incorporate with miniaturized optical systems or require complex fabrication techniques, and are sensitive to optical alignment. 

Here, we demonstrate simultaneous generation of concentric annular beams of different wavelengths using a one-dimensional photonic crystal cavity, which is a thin film based element that can be easily integrated with miniaturized optical systems and is robust to optical misalignment due to its planar architecture and translational invariance. Due to their compact architecture, photonic crystals are frequently employed for manipulating diffraction of output beams of micro-resonators \cite{Staliunas2007} and spatial dispersion of reflected beams \cite{Cheng2014}, and have been theoretically proposed for generating Bessel beams \cite{williams2005}. The photonic crystal cavity acts as a narrow bandpass filter in both spatial and temporal frequency domain \cite{liang2004}. We show that this filtering property, which is sensitive to the polarization direction of the incident light, combined with in-plane symmetry of the system transforms the fundamental Gaussian mode of an incident laser beam into an annular beam. We also show that when multiple laser beams of different wavelengths are coincident on the element, concentric annular beams of different diameters corresponding to different wavelengths are generated.

\section{\label{sec:level2}Methods}
\begin{figure*}[!]
\centering\includegraphics[width= 16 cm]{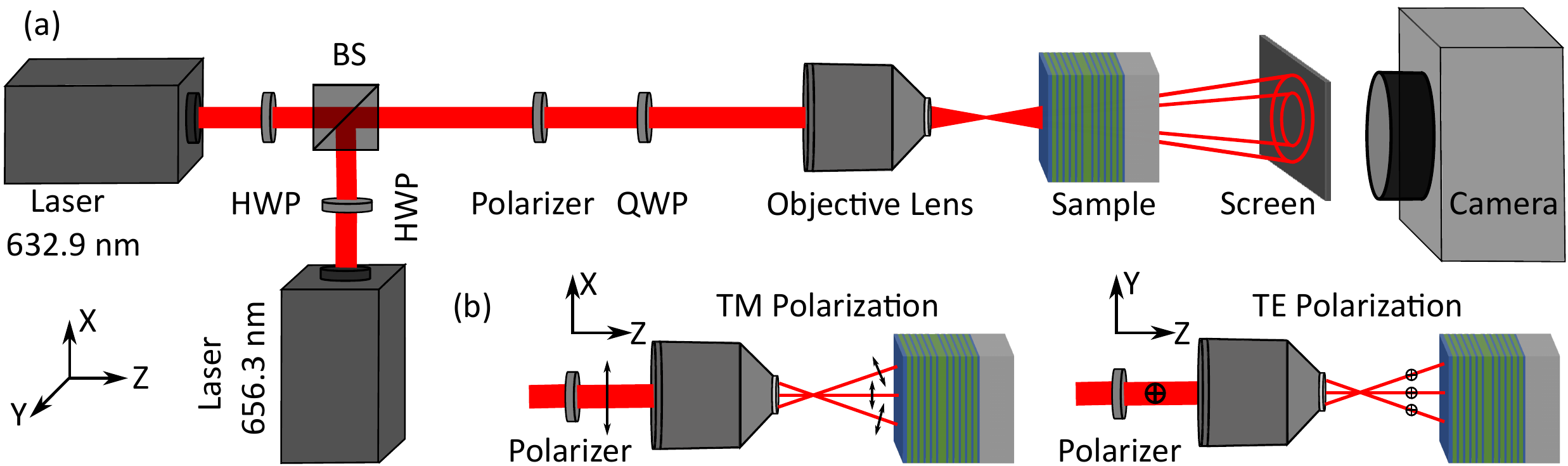}
\caption{\textbf{Experimental setup}: (a) Collimated Gaussian laser beams, at laser wavelengths of 632.9 nm and 656.3 nm, are focused using a 0.9 numerical aperture objective lens. A solid cone of wave-vectors is subtended on the fabricated sample and the transmitted beam is recorded using a screen and a camera. (b) Ray diagrams illustrating change in angle between linear polarization direction and plane of incidence of plane waves after passing through a high numerical aperture lens in two orthogonal planes. HWP: Half-wave plate. QWP: Quarter-wave Plate. BS: Beam splitter.} 
\vspace{-1.5em}
\label{Fig:Fig2_OpticalSetup}
\end{figure*}

We fabricate a photonic crystal cavity by introducing a silicon dioxide (SiO$_2$) defect between two distributed Bragg reflectors, each composed of six alternate thin films of SiO$_2$ and titanium dioxide (TiO$_2$)(Fig. \ref{Fig:Fig1_SEMandTransmission}a). We deposit all thin films using radio-frequency magnetron sputtering on a 1 inch$^2$ glass substrate. Experimentally measured transmission spectrum of the fabricated sample shows a stopband $\approx$ 560-770 nm and a cavity resonance at 672 nm (temporal quality factor $\approx$ 100) at normal incidence (Fig. \ref{Fig:Fig1_SEMandTransmission}b). These spectral features shift to lower wavelengths at oblique incidence of light \cite{wu2010,kumar2018}. We estimate the thicknesses of different thin films of the fabricated structure by fitting the experimentally observed transmission spectrum (Fig. \ref{Fig:Fig1_SEMandTransmission}b) using transfer matrix method \cite{Hecht2002} (defect layer: $\approx $ 250 nm, average thicknesses of SiO$_2$ and TiO$_2$ in Bragg reflectors: $\approx $ 110 nm and 80 nm with refractive indices 1.46 and 2.05, respectively).

We focus a linearly (or circularly) polarized and collimated Gaussian beam (at laser wavelengths 632.9 nm and 656.3 nm, optical power $\approx$ 200 $\mu W$, and beam diameter $\approx$ 1 mm; spatial profiles shown in Fig. A\ref{Fig:FigA1_LaserProfiles}) using a high numerical aperture (NA = 0.9) objective lens that subtends a solid cone of wave-vectors on the fabricated sample with maximum angle decided by the NA of the objective lens (Fig. \ref{Fig:Fig2_OpticalSetup}a).
A Gaussian beam can be considered as a distribution of plane waves with different propagation directions \cite{Landry1996}. A collimated beam, under paraxial approximation, has only one propagation direction and a defined polarization direction, say propagating along Z and linearly polarized along X (Fig. \ref{Fig:Fig2_OpticalSetup}b, before the objective lens when the quarter-wave plate is absent). As this beam is focused by a high NA lens, different plane waves bend towards the optic axis by different angles. This causes a change in the angle between the plane of incidence and the direction of polarization: for all the plane waves with plane of incidence as XZ plane, the polarization lies in the plane of incidence (Fig. \ref{Fig:Fig2_OpticalSetup}b, TM polarization), whereas for all the plane waves with plane of incidence as YZ plane, the polarization direction is perpendicular to the plane of incidence (Fig. \ref{Fig:Fig2_OpticalSetup}b, TE polarization). This results in different polarization directions for different parts of the beam focused by a high-NA objective lens \cite{Mansuripur1986}. We observe the transmitted beam at about 1 inch away from the sample using a screen and a camera (the sample and all components of the optical setup are mounted perpendicular to the propagation direction of the collimated laser beam). All images of the transmitted beam are adjusted to the same maximum brightness and the same contrast enhancement. 

\begin{figure*}
\centering\includegraphics[width=16 cm]{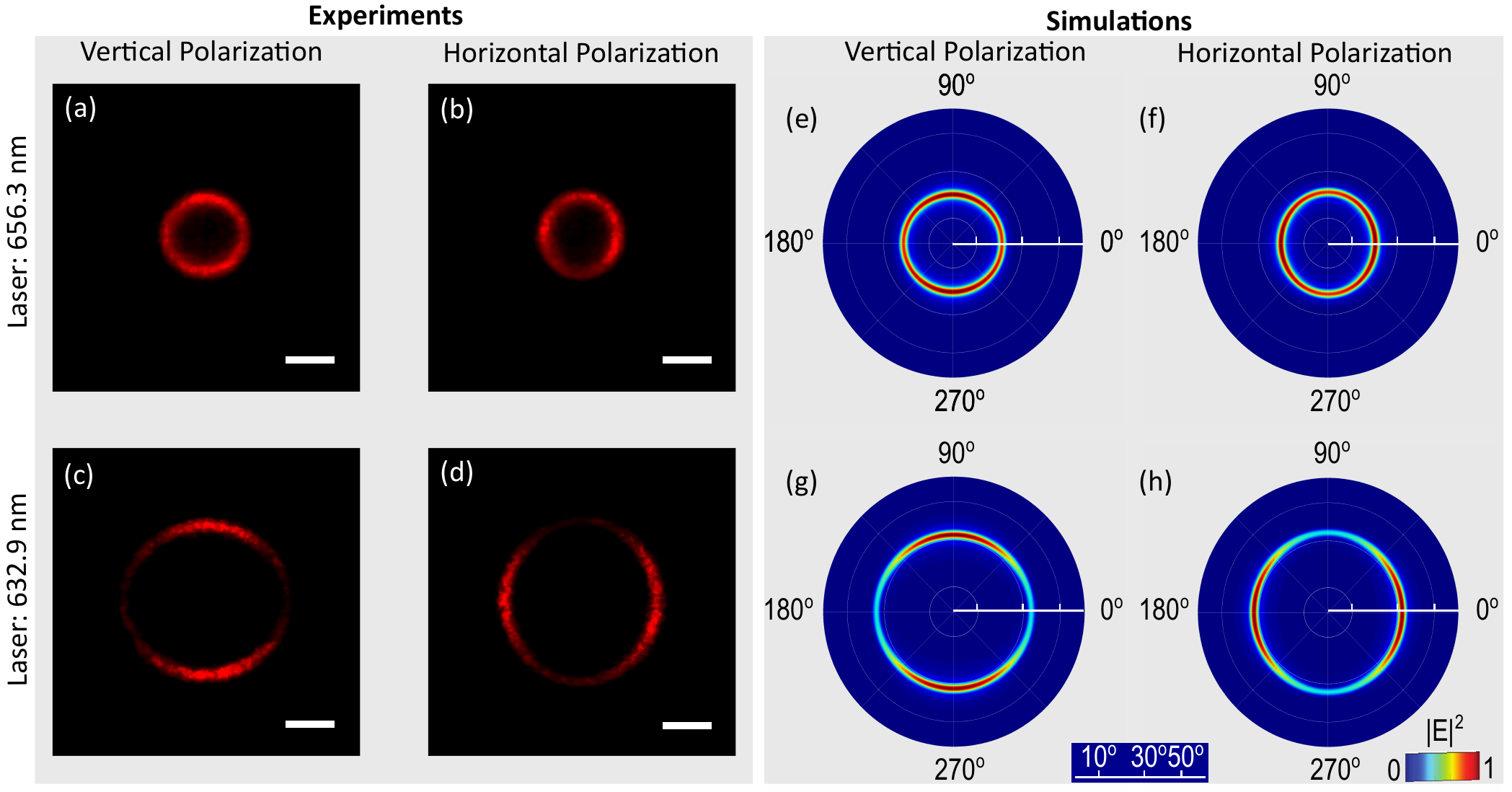}
\caption{\textbf{Transmitted intensity profiles for linearly polarized incident laser beams}:(a)-(d) Experimentally observed transmitted intensity profiles and (e)-(h) simulated far-field transmitted intensity ($|E|^2$ in $\theta-\phi$ axis system) profiles for two different wavelengths, 656.3 nm and 632.9 nm, and two orthogonal linear polarization directions (defined before the objective lens when the quarter-wave plate is absent) of the incident Gaussian beam at normal incidence. All scale bars are 10 mm.} 
\vspace{-1.5em}
\label{Fig:Fig3_Results_Linear}
\end{figure*}

\section{\label{sec:level3}Results}

\begin{figure}[t]
\centering\includegraphics[width=8cm]{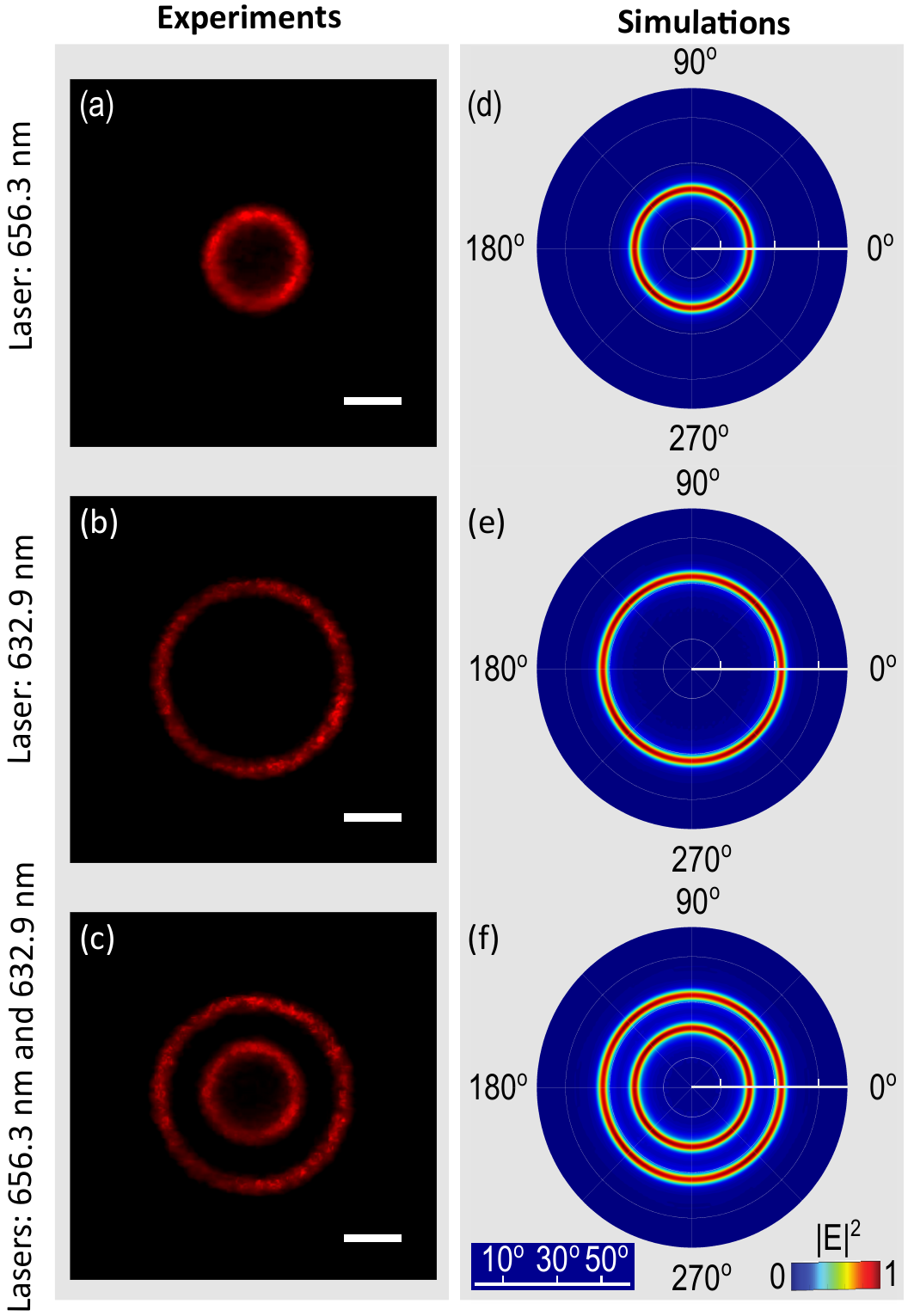}
\caption{\textbf{Transmitted intensity profiles for circularly polarized incident laser beams}: (a)-(c) Experimentally observed transmitted intensity profiles and (d)-(f) simulated far-field transmitted intensity ($|E|^2$ in $\theta-\phi$ axis system) profiles for laser wavelengths 656.3 nm, 632.9 nm, 656.3 nm and 632.9 nm combined, and circular polarization of the incident Gaussian beam at normal incidence. All scale bars are 10 mm.}
\vspace{-1.5em}
\label{Fig:Fig4_Results_Circular}
\end{figure}

We first measure the spatial profile of the transmitted beam for two orthogonal linear polarization directions of each of the two incident lasers at a fixed distance from the sample. For this measurement, quarter-wave plate (Fig. \ref{Fig:Fig2_OpticalSetup}a) is removed from the path of the incident beam. 
The incident laser wavelengths (656.3 nm and 632.9 nm) lie within the stopband of the photonic crystal (560-770 nm) and are less than the cavity resonance wavelength (672 nm), and therefore, they satisfy the resonance condition of the cavity at oblique incidence \cite{wu2010,kumar2018}. The high-NA objective lens before the sample generates a wide range of incidence angles simultaneously. Out of this large range of incidence angles, the cavity transmits only a narrow band of angles at the incident wavelengths, resulting into an annular spatial profile of the transmitted beam (Fig. \ref{Fig:Fig3_Results_Linear}a-d). We denote the center angle of this narrow band of transmitted angles as the resonance angle for the incident wavelength. The resonance angle is proportional to the difference between the incident wavelength and the cavity resonance wavelength at normal incidence. The transmitted beam is diverging away from the sample (divergence angle is equal to the resonance angle) because the focal point of the lens is between the lens and the sample. We observe that the diameter of the transmitted beam is larger for 632.9 nm laser than for 656.3 nm laser at the same distance from the sample. This difference in diameter is due to the larger resonance angle required for the 632.9 nm laser  than the 656.3 nm laser for cavity resonance wavelength of 672 nm at normal incidence. The width of the annular beam is determined by the spatial quality factor of the structure \cite{li1995}. 


We further observe that the annular beam has a smaller diameter in TM polarization than in TE polarization (Fig. \ref{Fig:Fig3_Results_Linear}a-d). By measuring diameter of the diverging beam at two different distances (17.4 mm apart) from the sample, we experimentally estimate the resonance angles to be 16.5$^{\circ}$ for 656.3 nm and 32.2$^{\circ}$ for 632.9 nm for TM polarization and 17.8$^{\circ}$ for 656.3 nm and 33.9$^{\circ}$ for 632.9 nm for TE polarization. Using transfer matrix method and plane-wave illumination, we confirm that indeed TM polarization has smaller resonance angle than TE polarization for the same incident wavelength, and therefore, has a smaller diameter. In general, we experimentally observe that the intensity of the transmitted beam changes with both polar and azimuthal angles. This angular variation in transmitted intensity is due to the presence of varied polarization directions in a beam focused by a high NA objective lens, which results in different transmission coefficients for different parts of the beam along polar and azimuthal directions. 

Using the estimated thicknesses of different thin films in the fabricated sample, we simulate the far-field transmitted intensity profiles of this photonic crystal cavity for a linearly polarized incident laser beam followed by a lens (NA = 0.9), using finite-difference time-domain (FDTD) simulations (Lumerical Inc.), as shown in figures \ref{Fig:Fig3_Results_Linear}e-h (radial direction represents polar angle). 
The resonance angle is larger for larger difference between cavity resonance wavelength and incident wavelength, as also observed in our experiments. In further agreement with the experiments, our simulations show that the resonance angle is smaller for TM polarization (19.9$^{\circ}$ for 656.3 nm and 32.3$^{\circ}$ for 632.9 nm) than for the TE polarization (20.9$^{\circ}$ for 656.3 nm and 33.8$^{\circ}$ for 632.9 nm) for the same incident laser. This anisotropy in polarization has also been observed in momentum space using a bare photonic crystal cavity at cryogenic temperatures \cite{Maragkou2011} and a bulky liquid crystal microcavity  \cite{Lekenta2018}, both in near-infrared part of spectrum, and explained using all-optical spin Hall effect. Using our FDTD simulations, we also calculate circular degree of polarization and linear degree of polarization \cite{Leyder2007} of the transmitted beam in momentum space (Fig. A\ref{Fig:FigA2_Degree_of_polarization}) that match well with observations made in experiments demonstrating all-optical spin Hall effect in bare photonic crystal cavities \cite{Maragkou2011, Lekenta2018}. Our simulations predict a spatial quality factor of about 10 for both polarizations at the respective resonance angles. Due to similar spatial quality factors and resonance angles for the two polarizations, the full width at half maximum of the annular beam is almost equal in the two orthogonal planes (Fig. \ref{Fig:Fig3_Results_Linear}e-h), though it appears to be different due to image-contrast in both experiments (Fig. \ref{Fig:Fig3_Results_Linear}a-d) and simulations (Fig. \ref{Fig:Fig3_Results_Linear}e-h). 

Next, we change the polarization of the incident laser beam from linear to circular by placing a quarter wave plate in between the polarizer and the objective lens (Fig. \ref{Fig:Fig2_OpticalSetup}a). The circular polarization ensures that both TE and TM components are equally present at every azimuthal angle after focusing by the high-NA objective lens. 
We observe circular annular beams with uniform intensity along azimuthal direction due to in-plane symmetry of the structure for circular polarization (Fig. \ref{Fig:Fig4_Results_Circular}a-b). We note that the angular separation between the TE and TM resonance is less than the resonance widths (Fig. \ref{Fig:Fig3_Results_Linear}e-h), therefore, under circular polarization, 
we observe only one annular beam in the transmitted beam. Using transmitted beam profiles captured 17.4  mm apart, we estimate that the average normalized width \cite{Lei2013} (defined as $1 - r_1/r_2$ where $r_1$ is the inner radius and $r_2$ is the outer radius) of the annular beam changes from 0.29 to 0.31 (standard deviation $\approx 7\%$) for 656.3 nm laser and from 0.12 to 0.15 (standard deviation $\approx 10\%$) for 632.9 nm laser as the beam propagates and diverges away from the sample. We attribute this slight increase in normalized width with propagation distance to reduction in signal to noise ratio as the beam diverges away from the sample, and can be potentially avoided by collimating the transmitted beam. Further, we generate two concentric annular beams of two different wavelengths by combining the two laser beams before the polarizer (Fig. \ref{Fig:Fig4_Results_Circular}c). We also spectrally characterize the concentric annular beams using a fiber-based spectrometer (Fig. A\ref{Fig:FigA3_Concentric_ring_Spectrometer}). This beam-shaping technique can be easily extended to multiple concentric annular beams by combining multiple sources with wavelengths smaller than the cavity resonance and lying within the stopband. We also simulate the far-field transmitted intensity profiles for this photonic crystal cavity for a circularly polarized incident laser beam using FDTD simulations. Our simulation results match well with our experimental observations (Fig. \ref{Fig:Fig4_Results_Circular}d-f). 

\section{\label{sec:level4}Conclusion and Discussion}
In conclusion, we demonstrated simultaneous generation of concentric annular beams of different wavelengths using a one-dimensional photonic crystal cavity. The photonic crystal cavity filters incident beams both spatially and spectrally. 
Using the in-plane symmetry of the structure to circular polarization, we demonstrated the generation of circular annular beams. The planar geometry and in-plane symmetry of the structure impart translational invariance, and thus, robustness to optical misalignment to our beam-shaping technique.
The total thickness of our structure is about 2 $\mu m$, and therefore, it can be easily deposited on output-facets of semiconductor and organic light sources \cite{Li2018} and fiber-optic devices \cite{Gissibl2016} for beam shaping without the need of an additional optical component. Multi-wavelength concentric annular beams can be employed for optical trapping of multiple types of nanoparticles \cite{Rodrigo2004} and for generating multi-wavelength Bessel beams. Our technique can also be extended for spectroscopy purposes as the resonance angle, and hence, the divergence of the annular beam is a function of the incident wavelength and its polarization.



\section*{Data Availability Statement}
\vspace{-1em}
The data that support the findings of this study are available from the corresponding author upon reasonable request.

\begin{acknowledgments}
\vspace{-1em}
SG acknowledges funding support from SERB (SB/S2/RJN-134/2014, EMR/2016/007113) and DST (SR/FST/ETII-072/2016). RV thanks IMPRINT-I (no. 4194) project for financial support. We thank Harshawardhan Wanare for insightful discussions. We acknowledge support of the FIB facility of MEMS department at IIT Bombay. 
\end{acknowledgments}

\appendix
\section*{Appendix}
\vspace{-1em}
\setcounter{figure}{0}
\begin{figure}[!ht]
\renewcommand{\figurename}{Fig. A}
\centering\includegraphics[width=7.5 cm]{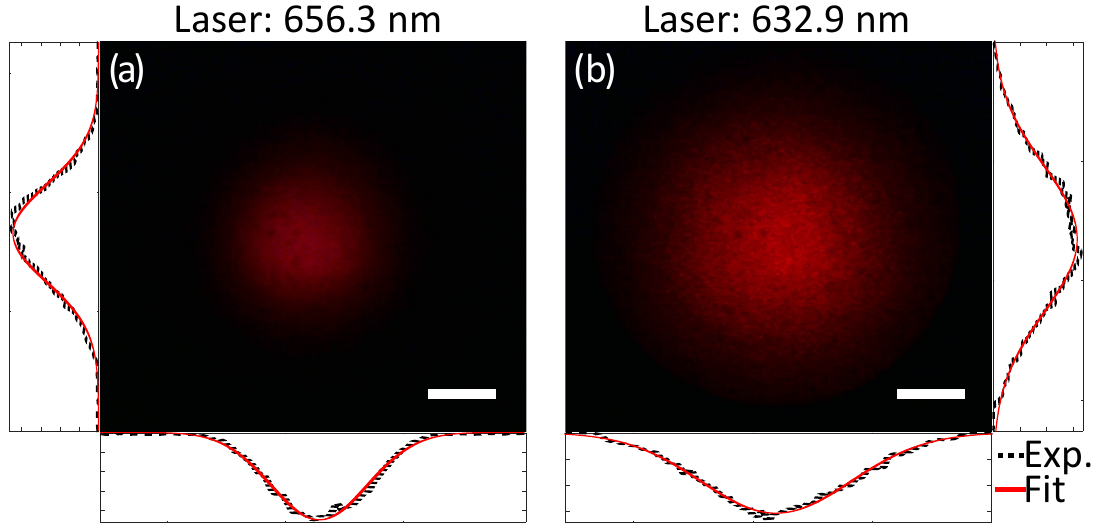}
\caption{\textbf{Spatial profiles of incident laser beams after the objective lens}: Measured intensity profiles of laser beams at wavelengths of (a) 656.3 nm and (b) 632.9 nm, along with Gaussian fits in two orthogonal directions. Intensity of lasers reduced to avoid saturation of camera. All scale bars are 10 mm.}
\label{Fig:FigA1_LaserProfiles}
\vspace{-2.5em}
\end{figure}

\begin{figure}[!b]
\renewcommand{\figurename}{Fig. A}
\centering\includegraphics[width=7.5 cm]{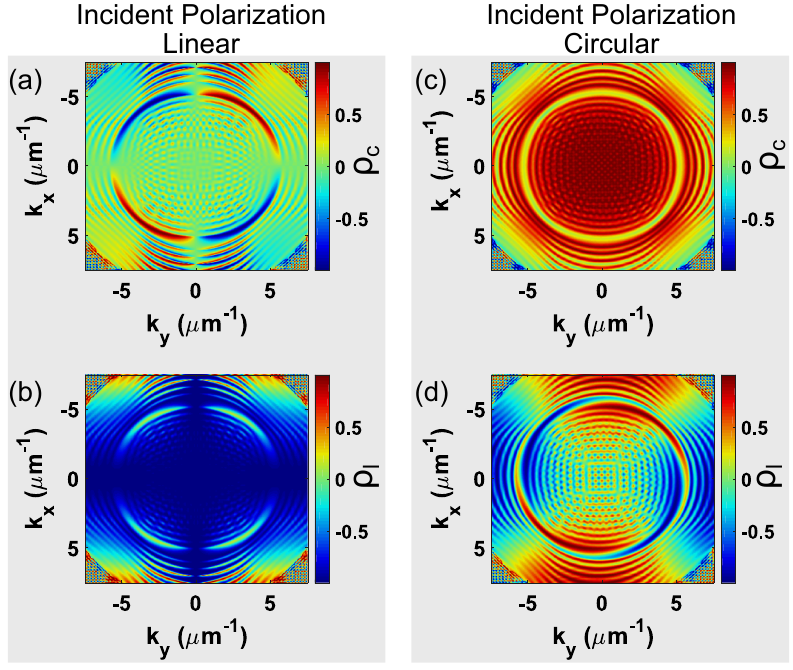}
\caption{\textbf{Degree of polarization}: Simulated degree of circular ($\rho_c$) and linear ($\rho_l$) polarization of the transmitted beam in momentum space for (a)-(b) linearly polarized incident laser beam, and (c)-(d) circularly polarized incident laser beam, at wavelength 632.9 nm.}
\label{Fig:FigA2_Degree_of_polarization}
\vspace{-2.5em}
\end{figure}

\begin{figure}[!h]
\renewcommand{\figurename}{Fig. A}
\centering\includegraphics[width=7.5 cm]{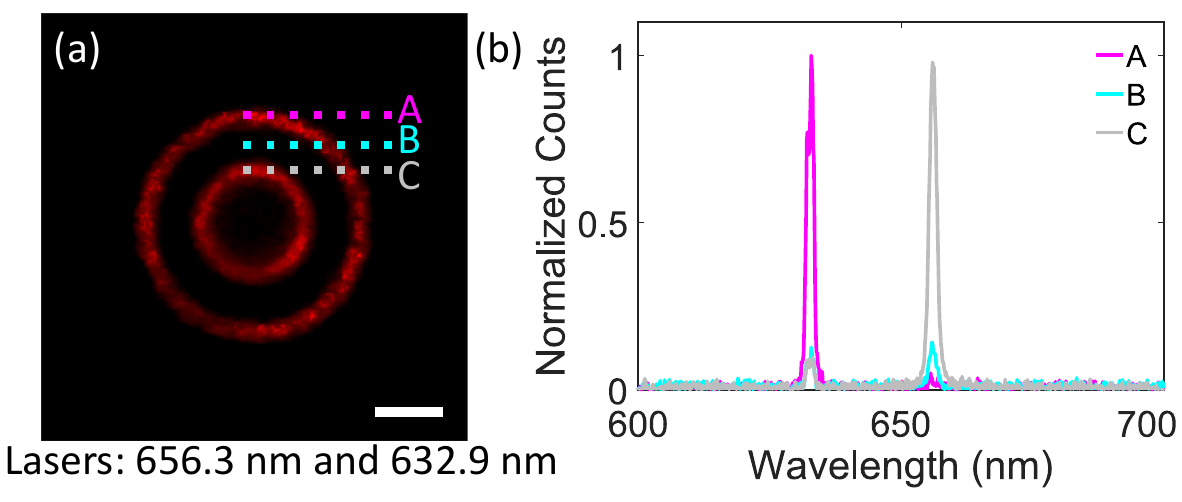}
\caption{\textbf{Spectral measurements on the transmitted beam}: (a)  Experimentally observed transmitted intensity profiles for simultaneously incident laser beams at wavelengths of 656.3 nm and 632.9 nm that are circularly polarized. (b) Spectra at three different locations in the output beam as marked in panel (a), using a fiber-based spectrometer (HR4000 Ocean Optics). The scale bar is 10mm.}
\label{Fig:FigA3_Concentric_ring_Spectrometer}
\end{figure}

\newpage
\section*{References}
\bibliography{references}
\end{document}